# Second-Order Nonlinear Microresonators on Thin-Film Lithium Tantalate

**Andrea Cordaro[1,*], Yunxiang Song[1,2], Keith Powell[1], Jiayu Yang[1], Shengyuan Lu[1], Donald Witt[1], Tong Ge[1], Xudong Li[1], Irene Occhiodori[1], Nicholas Achuthan[1], CJ Xin[1], Anna Shelton[1], Xinrui Zhu[1], Neil Sinclair[1], Marko Lončar[1,*]**

[1] John A. Paulson School of Engineering and Applied Sciences, Harvard University, Cambridge, MA, USA.

[2] Quantum Science and Engineering, Harvard University, Cambridge, MA 02138, USA

[*]e-mail: acordaro@g.harvard.edu; loncar@g.harvard.edu;

**Nonlinear frequency conversion enables coherent light generation in spectral regions that conventional lasers cannot reach directly and is foundational to the generation of squeezed states and correlated photon pairs. Implementing these nonlinear processes in microresonators lowers the required optical power by orders of magnitude but requires two conditions to be met simultaneously across wide optical bandwidths: phase matching between the interacting fields, and an appropriate coupling condition to the resonator both in the fundamental and second harmonic band. While the former can be addressed by quasi-phase matching through periodic poling, coupling two widely separated spectral bands to the resonator is nontrivial because of the wavelength dependence of mode overlap. Here, we demonstrate parametric oscillation and efficient second-harmonic generation in periodically poled racetracks on thin-film lithium tantalate on insulator. The proposed resonators feature bus-ring coupling engineered independently in the two spectral bands via a long synchronous directional coupler. Quasi-phase matching is achieved using sidewall poling that is also wavelength-calibrated and locally adapted to compensate for thickness nonuniformities. Together, these two design elements enable second-order parametric oscillation with an on-chip threshold of 181 μW and 9.8% conversion efficiency, with signal and idler tunable over tens of nanometers by a few hundred megahertz of pump detuning. Our design strategy can be flexibly applied to efficient second-harmonic generation, reaching 25% absolute conversion efficiency at milliwatt pump powers. These results show how the coupling condition is a design variable in second-order nonlinear resonators and establish thin-film lithium tantalate as a competitive platform for efficient and low-power frequency conversion, despite a second-order nonlinear coefficient roughly half that of lithium niobate.**

Second-order nonlinear optical processes have been key components for a wide range of photonic technologies since the earliest days of the laser[1,2]. Second-harmonic generation (SHG), sum- and difference-frequency generation (SFG, DFG), and optical parametric amplification and oscillation (OPA, OPO), now also demonstrated on photonic integrated platforms[3–6], provide coherent light in spectral regions where laser gain media are scarce, from the ultraviolet to the mid-infrared, by bridging it from well-developed optical bands. Moreover, these processes have many applications in quantum information and metrology, enabling the generation of squeezed states of light[7–10] and correlated photon pairs[11–16]. Finally, in the context of computing, networks of degenerate parametric oscillators were shown to implement Ising machines[17,18], while $\chi^{(2)}$ nonlinearities more broadly serve as the mixing and activation elements of all-optical neural and recurrent architectures operating at optical clock rates[19].

Integrated microresonators patterned directly in the nonlinear material dramatically lower the power required for these processes. Resonant enhancement of the circulating field at each of the interacting wavelengths, combined with the tight modal confinement available in integrated waveguides, has pushed parametric oscillation thresholds into the tens of microwatts range and brought second-harmonic absolute conversion efficiencies to tens of percent at milliwatt pump powers, all within compact footprints.[6,20–25]

In the past decade, established nonlinear bulk materials have become available as thin films on insulator, thus dramatically expanding the set of materials available for this purpose[26–29]. While thin-film lithium niobate on insulator (LNOI) has dominated integrated $\chi^{(2)}$ nanophotonics, Lithium tantalate on insulator (LTOI) has recently attracted attention as a closely related yet distinct platform. Historically, LTOI wafers have been manufactured at scale for radio-frequency filters, and more recently, low-loss waveguides, high-Q resonators, and electro-optic devices have been demonstrated[29–31]. However, compared to LNOI, lithium tantalate is known to exhibit a higher optical damage threshold and weaker photorefractive response[29,32–35]. Another convenient property of this platform is that lithium tantalate is nearly optically isotropic, with a material birefringence $\Delta n_{LT} \approx 0.004$, more than an order of magnitude smaller than in lithium niobate[29,36,37]. This, in turn, minimizes unwanted mode hybridization along waveguide bends[38], which, for example, has been recently shown to limit the tuning range of integrated lithium niobate parametric oscillators[6]. Importantly, the low birefringence of LTOI provides more flexibility in the choice of waveguide dimensions, so that modal purity does not need to be traded against confinement or dispersion engineering, also critical for efficient second-order nonlinear optics. Finally, periodic poling of thin-film lithium tantalate has been recently achieved[32,39–41], albeit with a nonlinear coefficient about half that of lithium niobate[42]. Realizing $\chi^{(2)}$ microresonators in integrated photonics requires simultaneous control of two conditions across an octave of optical bandwidth. First, phase mismatch between the fundamental and second harmonic needs to be addressed by quasi-phase matching via periodic poling. Second, the coupling of each band between the resonator and the bus waveguide needs to be optimized. This is a non-trivial task because the evanescent overlap between bus and resonator falls off steeply with frequency, so a coupler that is critically coupled in the telecom band extracts almost nothing in the visible, and one optimized for the visible does not guarantee optimal coupling at telecom. While poled racetracks in LTOI have very recently been demonstrated for photon pair generation via spontaneous parametric down-conversion[43], parametric oscillation has not been demonstrated on this platform in any geometry.

Here we demonstrate second-order nonlinear microresonators in thin-film lithium tantalate in which the coupling to the bus waveguide is tailored independently in octave-separated wavelength ranges, and report, to our knowledge, the first parametric oscillation on this platform, as well as efficient second-harmonic generation. Our periodically poled racetracks incorporate a long directional coupler in which the bus and resonator waveguides have the same width, and therefore the same effective index, so that light transfers completely back and forth between them along the coupling region. This, in turn, means the per-pass coupling can be arbitrarily set between zero and close to unity and can be positioned in wavelength independently at the fundamental and second harmonic. Quasi-phase matching is provided by periodic ferroelectric domain inversion. Specifically, we performed sidewall poling, which has been shown to provide consistent and complete domain inversion[5] and, to our knowledge, has not previously been demonstrated in lithium tantalate. We combined this technique with wavelength-calibrated adaptive poling to align the quasi-phase-matching transfer function with the designed coupling behavior and compensate for film-thickness variation across the device. Using appropriately designed microresonators, we demonstrate parametric oscillation with an on-chip threshold of ~181 μW and ~9.8% pump-to-signal-and-idler conversion, with the oscillating pair tunable across more than 30 nm by a few hundred megahertz of pump detuning. Next, we show efficient second harmonic generation yielding

~25% absolute conversion efficiency at milliwatt pump powers. Therefore, the design strategy proposed here can be flexibly applied to two opposite processes (OPO, SHG) by tailoring the bus-ring coupling spectrum relative to the degenerate quasi-phase-matching wavelength.

**Results**

The nonlinear resonators studied in this work are periodically poled racetracks in thin-film lithium tantalate that involve fundamental-harmonic (FH) modes at telecom near 1560 nm and second-harmonic (SH) modes in the visible near 780 nm. The intrinsic $\chi^{(2)}$ nonlinearity of the material couples two FH resonances at angular frequencies $\omega_m$ and $\omega_n$ to an SH resonance at $\Omega_k$, with energy conservation fixing $\omega_m + \omega_n = \Omega_k$. For second-harmonic generation (SHG), the two FH photons are drawn from a single mode (m = n, $2\omega_m = \Omega_k$), while optical parametric oscillation runs the same interaction in reverse, splitting one SH pump photon ($\Omega_k$) into a signal and idler pair ($\omega_m, \omega_n$). Leveraging high quality factor resonances in both bands extends the photon lifetimes of the interacting modes, thereby strongly enhancing the nonlinear interaction and hence the effective conversion, which scales with the quality factors of the participating modes. Our microresonators consist of a racetrack of two 2 mm straight sections joined by a pair of Euler bends (r = 200 µm), giving a round-trip length of about 5.6 mm at a waveguide width of 2 µm (Fig. 1a). The devices are fabricated on x-cut lithium tantalate, for which the crystal z-axis lies in the chip plane (see Methods). The straight sections are oriented with their propagation direction perpendicular to z. The design of these resonators for both SHG and OPO processes is essentially set by two requirements: efficient pumping and extraction of light across widely separated bands, achieved with an engineered bus-ring coupler, and phase matching between the FH and SH, achieved by periodic poling.

Designing a coupler that is simultaneously efficient at both telecom and visible wavelengths is challenging because the phase matching and the spatial overlap between the bus and ring modes both depend strongly on wavelength, with the overlap decaying as wavelength decreases. To gain control over coupling across these far-separated bands, we employ a long directional coupler whose operating principle is inspired by the pulley couplers used in microcomb resonators[44,45]. We compute the per-pass power coupling coefficient from bus to ring, $\mathrm{K_{pass}}$ ($TE_{00}$ to $TE_{00}$ in both bands), using full-wave FDTD simulations. FDTD is preferred here over spatial coupled-mode theory (CMT) because it captures the regions where the bus is gradually brought towards and away from the ring, which are harder to capture with CMT. From $\mathrm{K_{pass}}$ we obtain the extrinsic coupling rate $\kappa_e$ and the extrinsic quality factor $\mathrm{Q_e} = \frac{\omega}{\kappa_e} = \frac{\omega}{\mathrm{K_{pass}\,FSR}}$, where $FSR$ is the free spectral range (~22.8 GHz at FH and ~22.2 GHz at SH).

The coupler consists of a 0.7 µm wide bus that is brought close to the bottom straight of the ring through two Euler bends (15°, r = 120 µm); over the coupling region, the racetrack's bottom straight section is tapered to 0.7 µm, and the two run parallel over a length L at a top gap of 550 nm (excluding the slanted sidewalls). Figure 1b shows the simulated field profiles at the SH (visible) and FH (telecom) wavelengths. The much stronger overlap at telecom means that light couples into and back out of the ring over the coupler, leaving just the right amount in the ring, while at the visible wavelength the weaker overlap is still sufficient to reach critical coupling. Next, from the extrinsic quality factor, it is possible to calculate the coupling efficiency (or extraction efficiency) defined as $\eta = \frac{Q_i}{Q_i+Q_e}$ , which we evaluate assuming an intrinsic quality factor $\mathrm{Q_i}$ = 2 M, close to the experimental values (see Methods). By adjusting the gap and the length L, we can hold the visible near critical coupling while placing the telecom anywhere from strongly overcoupled ($\mathrm{K_{pass}} \rightarrow 1$ so that $\eta \rightarrow 1/(1 + (\omega/\mathrm{FSR})/\mathrm{Q_i}) \approx 1$), to fully uncoupled, $\mathrm{K_{pass}} \rightarrow 0$ so that $\eta \rightarrow 0$. Figure 1c shows the simulated coupling efficiency $\eta$ as the coupler length L is swept. In the visible, the curves cluster near $\eta = 0.5$ (critical coupling), while at telecom $\eta$ varies strongly with wavelength, dropping to sharp nulls and rising toward unity (overcoupled) in between. Because the bus and ring are the same width (0.7 μm) over the coupling region, power transfers fully back and forth between them along the interaction length. Nulls occur where a complete number of transfer cycles returns all light to the bus. Since the amount of coupling per unit length grows towards longer wavelengths as the modes become less confined, longer L shifts the zero-net-coupling point to a shorter wavelength. Toward smaller wavelengths, the coupling efficiency drops due to mode overlap decay.

The other key requirement is phase matching between FH and SH. Indeed, due to material and geometric dispersion, the FH and SH modes have different phase velocities, thus hindering efficient nonlinear conversion. The momentum mismatch $\Delta\mathrm{k} = (\mathrm{n_{eff,k}}\Omega_\mathrm{k} - \mathrm{n_{eff,m}}\omega_\mathrm{m} - \mathrm{n_{eff,n}}\omega_\mathrm{n})/\mathrm{c}$ is compensated by quasi-phase-matching (QPM): the sign of $\chi^{(2)}$ is reversed periodically by inverting the ferroelectric domains, supplying the missing momentum through a poling grating of period $\Lambda = 2\pi/\Delta\mathrm{k}$. We take advantage of the largest component of the $\chi^{(2)}$ tensor by using the fundamental TE mode at both FH and SH, given the orientation of our device. Because the phase-matching condition depends strongly on film thickness, which varies across the chip on scales comparable to our device size, we use wavelength-calibrated adaptive poling[46,47], adjusting the poling period along the resonator to correct for thickness variation and to align the degenerate phase-matching wavelength ($\Delta\mathrm{k} - 2\pi/\Lambda = 0$ for $\omega_\mathrm{m} = \omega_\mathrm{n} = \Omega_\mathrm{k}/2$) with the spectral response of the coupler. Domain inversion is achieved by sidewall poling, to our knowledge not previously demonstrated in lithium tantalate, in which the poling electrodes are placed on the sides of the waveguide. Figure 1d shows an SEM of these electrodes, confirming good alignment to the waveguide, set by aligned e-beam lithography with the aid of local markers. The SHG confocal microscopy image in Fig. 1e shows reliable domain inversion with a duty cycle close to 50%.

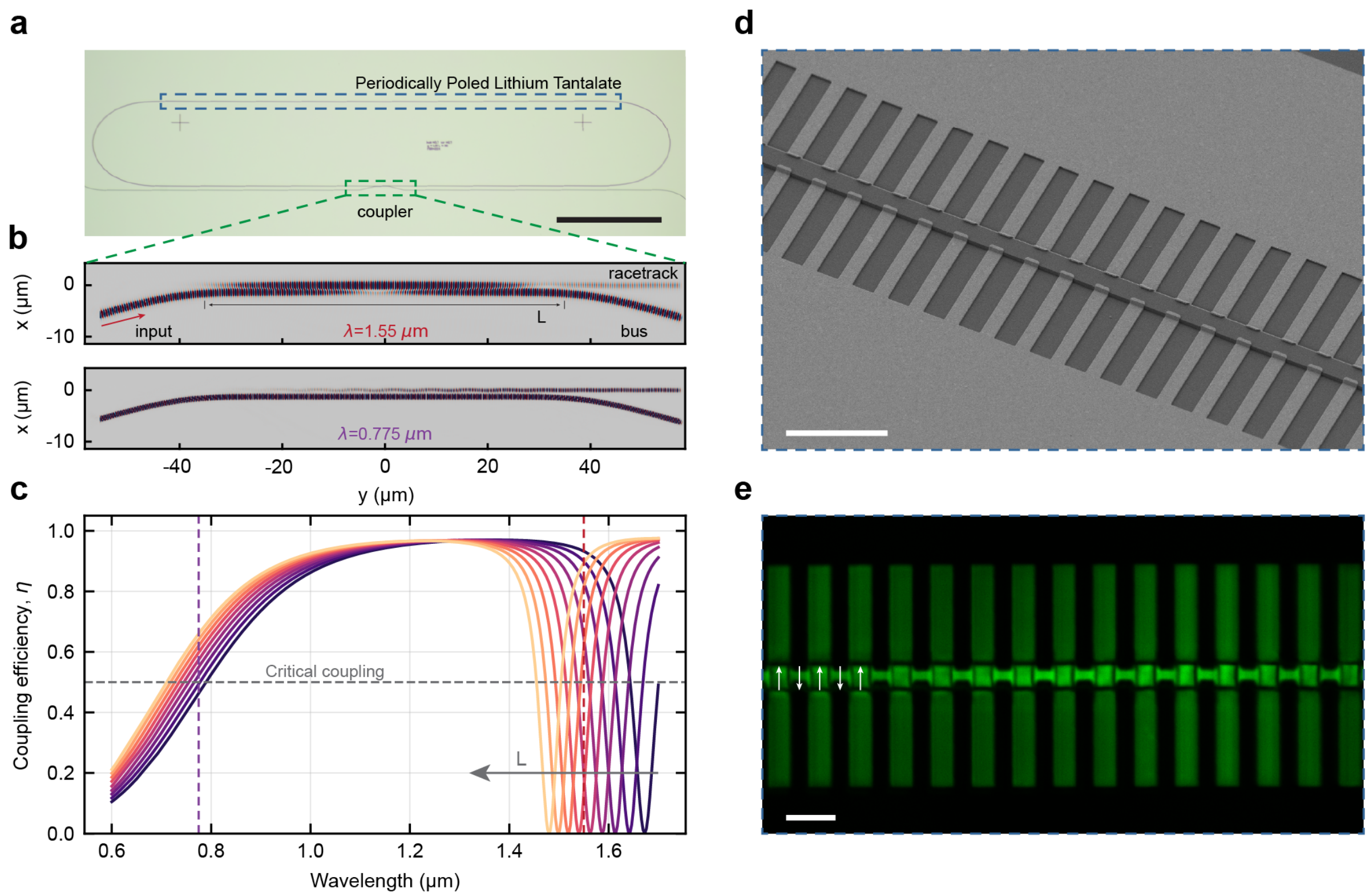

Fig. 1 | **Periodically poled LT microresonator. a.** Optical microscope image showing the microresonator geometry and indicating the periodically poled region and the coupler region (scale bar 500 µm). **b**. Simulated field profiles (real part of $E_z$) at $\lambda$ =775nm and $\lambda$=1550 nm for a coupler geometry that is close to critically coupled at both wavelengths (gap=550nm, coupler length L=47 µm). **c.** Simulated coupling efficiency $\eta$ as the coupler length L is swept from 30 µm (dark purple line) to 54 µm (orange line) in 9 steps. The dashed gray horizontal line ($\eta$ = 0.5) indicates critical coupling. The two vertical dashed lines indicate the design wavelength at the FH and SH spectral bands. **d**. SEM image of the resonator top straight section with deposited poling electrodes (scale bar 3 µm). **e**. Confocal second-harmonic microscopy image (scale bar 5 µm) showing the inverted domains between the poling electrodes (black shadows). White arrows indicate the orientation of the crystal +z axis, which alternates between adjacent domains.

Combining the two design knobs described above, the wideband coupler and quasi-phase-matched poling, we demonstrate optical parametric oscillation (Fig. 2a). For this device, we set the degenerate poling phase-matching wavelength to coincide with the coupler's telecom coupling null, so that the degenerate point, where signal and idler would meet, sits where the ring is decoupled from the bus. The OPO runs nondegenerate, hence energy conservation places signal and idler symmetrically in frequency about this point, and as they separate from degeneracy, each moves out of the coupling null toward critical coupling, so both are extracted efficiently. Figure 2b shows the linear telecom transmission spectrum. The coupler's wavelength dependence is directly visible: near 1560 nm the ring resonances nearly disappear, since no light couples into the ring at the $\eta = 0$ wavelength, while away from this condition the resonances deepen as coupling grows. On the other hand, the visible (SH) transmission spectrum shows resonances that are close to critical coupling, confirming that the SH pump is efficiently loaded into the ring (see Fig. 2c). Figure 2d shows the OPO output as the pump is tuned across the same range, demonstrating many lines above threshold. The generated telecom light is separated from the residual pump with a WDM (see Methods), and the plotted signal is the collected telecom power, so that oscillation appears wherever the pump is resonant and above threshold.

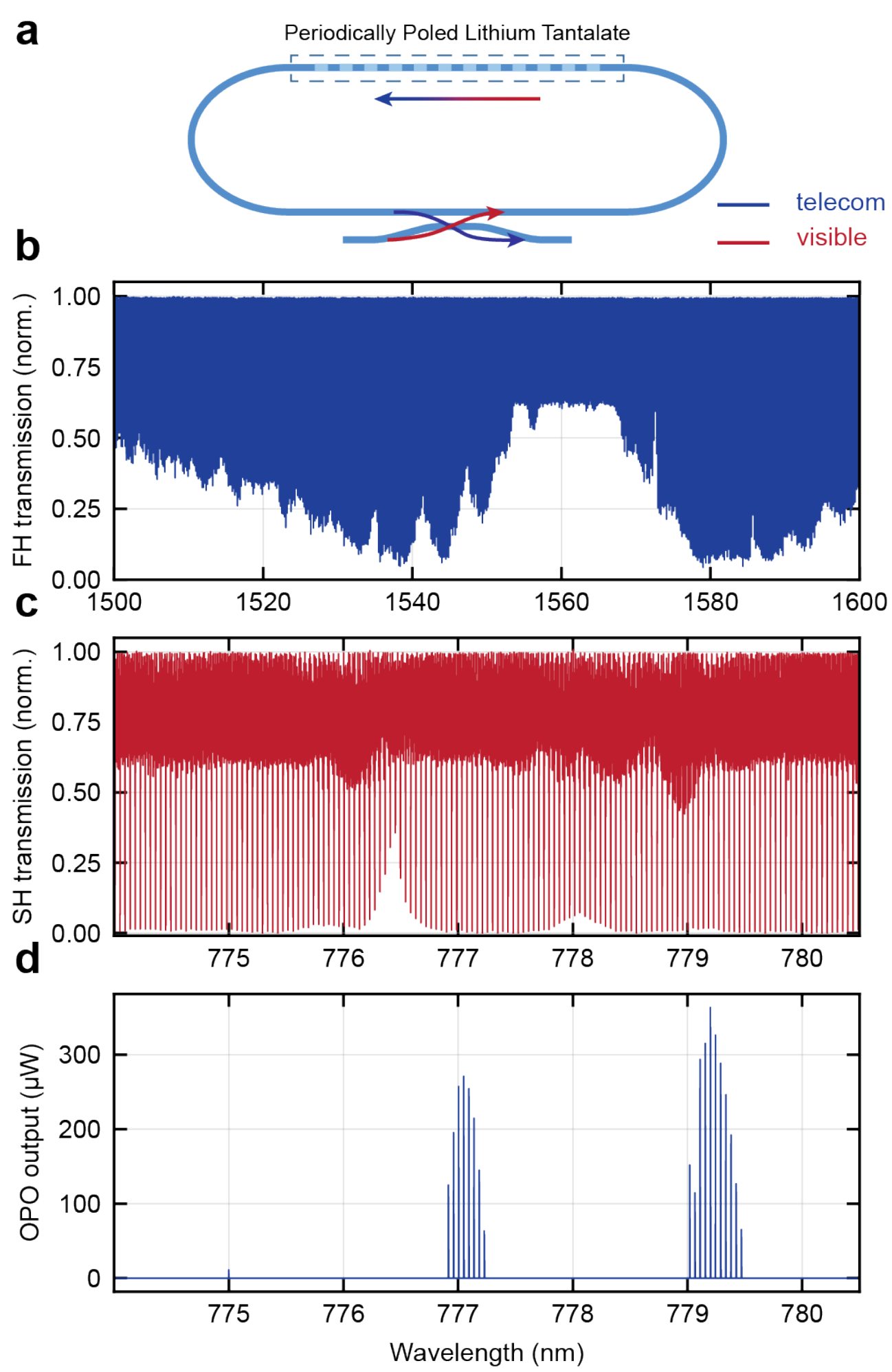

Fig. 2 | **Optical parametric oscillation. a.** Schematic of the device: the visible SH pump is coupled into the racetrack through the bus, and the generated telecom signal and idler are collected at the output. The periodically poled section is indicated. **b.** Normalized FH (telecom) transmission spectrum, showing the strong wavelength dependence of the coupler; the ring resonances nearly vanish around 1560 nm, where the bus and ring have near-zero coupling, and deepen away from it as coupling increases. **c**. Normalized second-harmonic transmission spectrum, with resonances close to critical coupling, confirming efficient loading of the SH pump. **d.** On-chip OPO output as the SH pump is scanned across the same visible range. The generated telecom light is separated from the residual pump using a wavelength-division multiplexer, and its power is recorded.

To quantify the parametric oscillation, we model the device using the standard three-wave mixing $\chi^{(2)}$ Hamiltonian, in which one second-harmonic pump mode at $\Omega_0$ is coupled to a signal-idler pair at $\omega_{\pm m} = \omega_0 \pm D_1 m + \frac{1}{2} D_2 m^2$ (where $D_1 = 2\pi \times FSR$ and $D_2$ is the second-order integrated dispersion) via the nonlinear rate $g_{0,-mm}$ . From the Hamiltonian and the related equations of motion for the FH and SH field amplitudes, it is possible to derive the visible pump threshold power[22]

$$\mathrm{P_{th}} = \frac{\hbar\Omega_0}{16\left|g_{0,-mm}\right|^2 \kappa_\mathrm{B}^{(e)}}\left(\Delta^2 + \frac{\kappa_\mathrm{B}^2}{4}\right)\left[(\Delta + \mu - \mathrm{D_2 m^2})^2 + \kappa_{(\mathrm{A,m})}\kappa_{(\mathrm{A,-m})}\right] \quad (1)$$

where $\Delta = \Omega_\mathrm{L} - \Omega_0$ is the pump laser detuning, $\mu = \Omega_0 - 2\omega_0$ is the modal detuning between the second-harmonic and fundamental resonance ladders, $\kappa_{(\mathrm{i,j})}$ is the total cavity loss rate of mode j on the fundamental (i=A) or second-harmonic (i=B) ladder. Intuitively, the threshold power decreases for smaller linewidths both at telecom and visible and higher nonlinear coupling rate, while it increases with detunings. The term $(\Delta + \mu - \mathrm{D_2 m^2})$ is the doubly resonant detuning: the frequency mismatch between the signal-idler pair fixed by energy conservation and the fundamental FH cavity resonances on which they must oscillate. It vanishes for the mode pair that simultaneously satisfies energy conservation and double resonance, and the threshold is minimized there. For a given pump frequency, the signal-idler pair that has the lowest $\mathrm{P_{th}}$ will oscillate first. Above the oscillation threshold, the intracavity pump field clamps at its threshold value, and the generated fundamental power follows the characteristic square-root law

$$\mathrm{P_{out}} = \frac{4\eta_\mathrm{B}}{\Omega_0}\left(\eta_{(\mathrm{A,m})}\omega_\mathrm{m} + \eta_{(\mathrm{A,-m})}\omega_{(\mathrm{-m})}\right)\mathrm{P_{th}}\left(\sqrt{\mathrm{P_{in}/P_{th}}} - 1\right) \quad (2)$$

where $\mathrm{P_{in}}$ is the SH pump power, and $\eta_\mathrm{A}$, $\eta_\mathrm{B}$ are the out-coupling (extraction) efficiencies of the fundamental and second-harmonic modes.

Figure 3 shows the measured OPO output power for a specific pair as a function of the SH pump power (on-chip powers, corrected for coupling loss) while driving a resonance around 779.14 nm. The fundamental output power (Fig. 3a) exhibits the square-root turn-on of Eq. (2), and a fit of the same equation yields a threshold of $\mathrm{P_{th}}$= 181 µW (see Methods). The extracted conversion efficiency (Fig. 3b) rises steeply above threshold, reaches a maximum of about 9.8% near 832 µW, and then declines slowly following the expected pump-depletion behavior. Indeed, the overall pump-to-signal/idler conversion efficiency is proportional to the pump depletion $\mathrm{D} = 4\sqrt{\mathrm{P_{th}/P_{in}}} - 4\mathrm{P_{th}/P_{in}}$ , which rises from zero at threshold, reaches a maximum at $\mathrm{P_{in}}$ = $4\mathrm{P_{th}}$, and slowly decreases at higher pump power. The experimentally measured peak position and rolloff are in good agreement with this dependence.

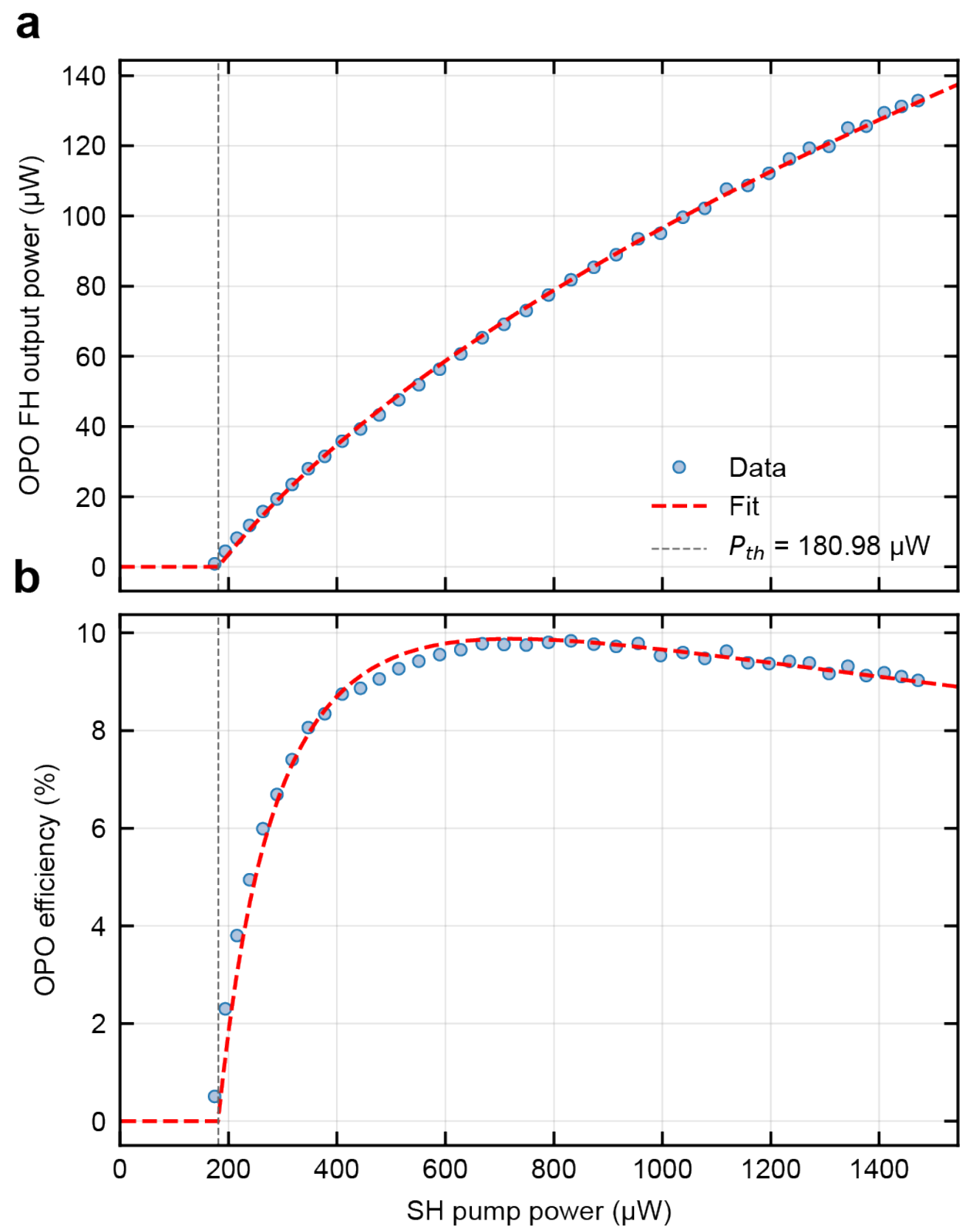

Fig. 3 | **OPO threshold and conversion efficiency. a.** FH (telecom) on-chip output power versus SH (visible) on-chip pump power. The output shows the characteristic square-root turn-on above the oscillation threshold; a fit (dashed red line) gives a threshold of $P_{th} \approx$ 181 μW. **b.** OPO conversion efficiency (generated telecom power divided by SH pump power) over the same SH pump power range as in panel a. The efficiency rises steeply above threshold, reaches a maximum of 9.8% near 832 μW (experimental data), and declines gradually at higher power. Dashed lines are fits.

To study the spectral content of the OPO, we drive the device at the same SH resonance (779.14 nm) and send part of the telecom output to an optical spectrum analyzer (OSA). The doubly-resonant condition fixes the signal and idler wavelengths through the pump and modal detunings: changing Δ or μ selects a different mode m that minimizes the term $(\Delta + \mu - D_2 m^2)$ in Eq. (1). At fixed operating temperature, the oscillating pair can therefore be tuned by the pump detuning alone (Fig. 4a). As the pump frequency is scanned across the SH resonance via the laser's internal piezo (detuning Δ), the parametric gain favors successive fundamental mode pairs, and the signal and idler shift symmetrically away from the degenerate point. Figure 4b maps the telecom output versus Δ. The signal and idler branches diverge from degeneracy and separate by tens of nanometers over a pump detuning of only a few hundred megahertz. Figure 4c shows the second-harmonic transmission and total fundamental output recorded concurrently during the same scans. The intensity steps in the output mark the transitions from one oscillating mode pair to the next.

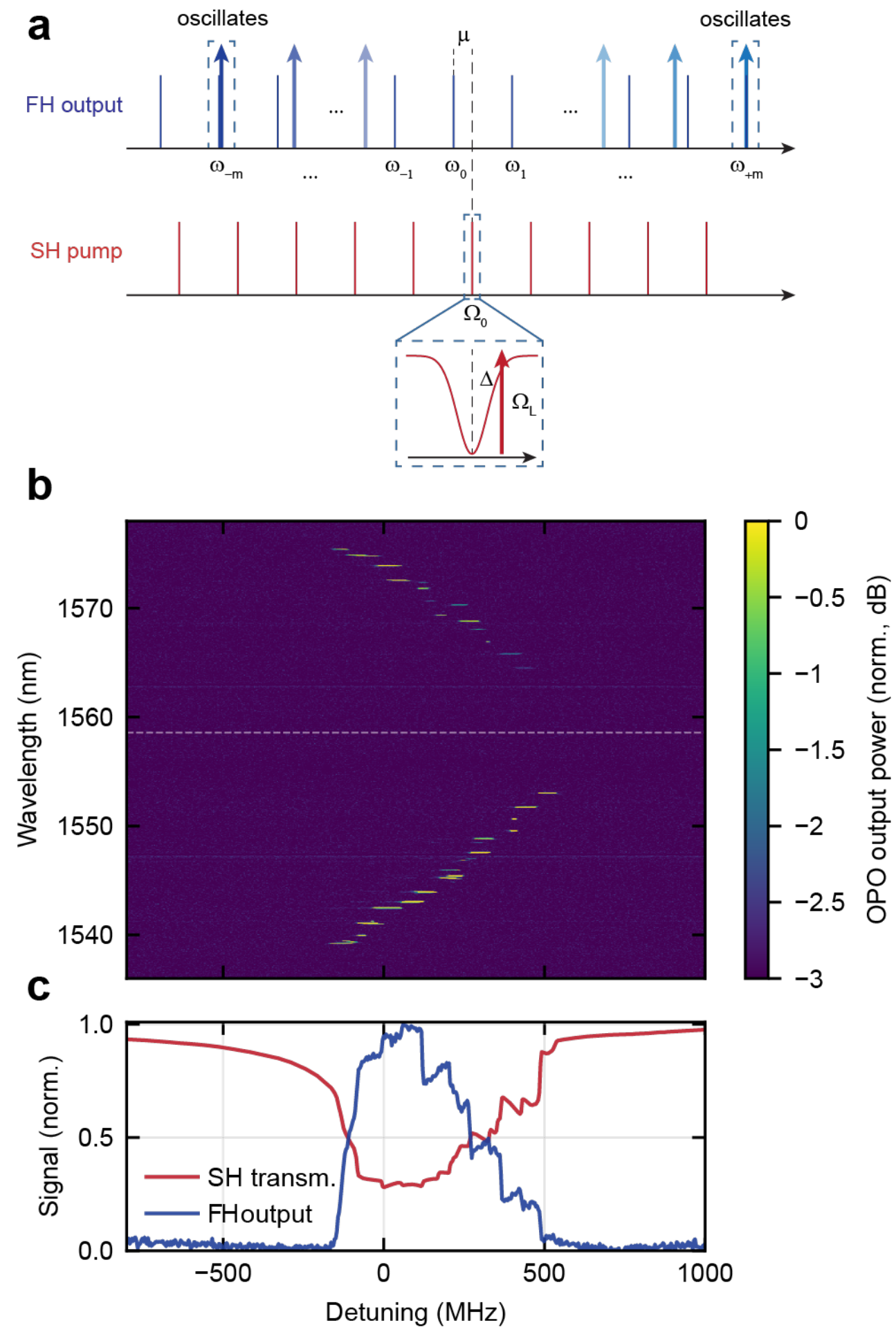

Fig. 4 | **OPO output spectral control. a.** Schematic of the process. The second-harmonic pump, sitting on one SH resonance, drives a signal-idler pair placed symmetrically in frequency about the modal-detuning point μ; the intracavity pump detuning Δ within the SH resonance is shown in the inset. Tuning Δ causes the wavelengths of the oscillating pair to be constrained to successive fundamental mode pairs. **b.** Fundamental harmonic (telecom) output as a function of SH pump detuning Δ across a single SH resonance. As Δ is tuned, the signal and idler move away from the degenerate point near 1558.28 nm (dashed line), spanning ∼36 nm over a few hundred megahertz of pump detuning. Color is normalized output power (dB). **c.** SH transmission (red) and total fundamental output (blue) versus Δ over the same resonance.

The same coupler concept can be used to optimize second-harmonic generation by making the opposite design choice. We use a second device in which the poling is set so that the phase-matched fundamental band lies in the overcoupled region of the coupler, rather than at the coupling null. For this ring, the fundamental mode is strongly overcoupled at low power, with an extraction efficiency $\eta_A \approx 0.84$, so that its resonances are shallow (Fig. 5a). As the telecom pump power increases, the second-harmonic conversion acts as an additional, power-dependent loss on the fundamental mode. The nonlinear loss adds to the intrinsic loss, bringing the total internal loss rate closer to the external coupling rate, and the phase-matched resonances get closer to critical coupling, where pump light is most efficiently loaded into the ring, and second-harmonic generation is maximized. This effect is visible in the deepening of the phase-matched resonances at high pump power (Fig. 5b). Figure 5c shows the second-harmonic output over the same fundamental-wavelength scan. The SHG peak positions are set by the doubly-resonant modes selected by the FH and SH free-spectral-range mismatch (a Vernier condition), while the envelope follows the quasi-phase-matching ($\text{sinc}^2$ function) response of the poling.

To quantify the conversion, we drive a single phase-matched resonance near 1559.89 nm and record the second-harmonic power as a function of on-chip fundamental pump power (Fig. 5d). We model the response by solving the coupled FH and SH cavity equations of motion for the intracavity fundamental field and the resulting second-harmonic output (see Methods). At very low power, the SHG output follows the expected quadratic dependence, so the efficiency rises linearly with pump power (Fig. 5e). The efficiency curve deviates quickly from a linear behavior and reaches a maximum of about 25% near a telecom input power of 3.4 mW and falls off at higher power. Indeed, at this pump power, the intracavity SH field becomes large enough to parametrically generate new telecom sidebands, i.e., a cascaded OPO. The onset of cascaded oscillation is confirmed by the appearance of new telecom lines in the output spectrum (Fig. 5e, inset).

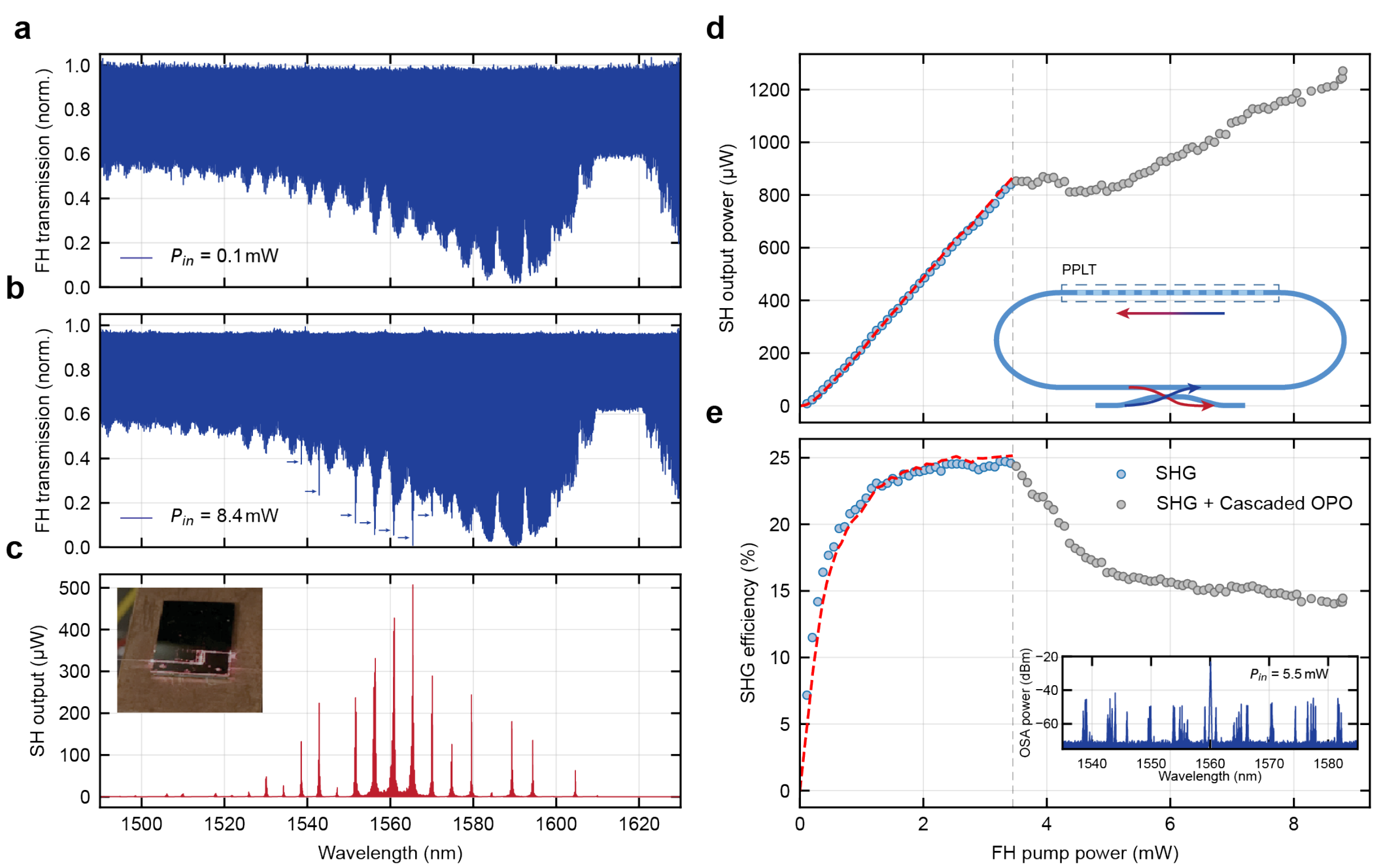

Fig. 5 | **Second-harmonic generation with nonlinear critical coupling**. **a.** Normalized FH (telecom) transmission spectrum at low pump power ($P_{in}$ = 0.1 mW); the ring resonances are shallow because the fundamental mode (1559.89 nm) is strongly overcoupled ($\eta_A \approx$ 0.84). **b.** Normalized FH transmission spectrum at high pump power ($P_{in}$ = 8.4 mW). The phase-matched resonances indicated by blue arrows deepen as the power-dependent nonlinear loss brings them closer to critical coupling. **c** SH output

versus fundamental pump wavelength. Inset: photograph of the fiber-coupled device. **d.** On-chip SH power versus on-chip fundamental pump power for a single phase-matched resonance near 1559.89 nm. Blue points are the data used in the fit (below cascaded OPO threshold), the red dashed curve is the model, and the gray points lie above the cascaded-OPO onset (dashed vertical line) and are not described by the two-modes model. Inset: device schematic. **e.** Corresponding SHG efficiency, reaching about 25%. Inset: output spectrum at $P_{in}$ = 5.5 mW, showing the telecom lines generated by the cascaded OPO. The spectrum is taken by averaging ten OSA scans while piezo-scanning the telecom laser around the FH resonance.

## Conclusion

To conclude, we have demonstrated second-order nonlinear microresonators in thin-film lithium tantalate, realizing both optical parametric oscillation and second-harmonic generation in periodically poled racetracks. Two design elements make this possible. The first is a long phase-matched directional coupler that provides independent control of the coupling at telecom and visible wavelengths, circumventing the frequency-dependent overlap that ordinarily prevents a single coupler from serving both bands. The second is quasi-phase-matching by wavelength-calibrated adaptive sidewall poling that compensates for film-thickness variation across the device. Placing the quasi-phase-matching point at the coupler's zero-net-coupling wavelength yields parametric oscillation with an on-chip threshold of 181 µW and a pump-to-signal-idler conversion efficiency of 9.8%, with signal and idler tunable across tens of nanometers by a few hundred megahertz of pump detuning. Making the opposite design choice and placing the phase-matched fundamental in the overcoupled region yields 25% second-harmonic conversion efficiency at a few milliwatt-level pump powers. Overcoupling the visible mode, which the coupler design permits independently of the telecom band, could raise the conversion efficiency even higher, as the threshold for cascaded OPO is also pushed towards higher powers.

Further extensions of this work may include dispersion engineering to achieve broader phase matching. This typically pushes waveguide dimensions toward thicker films, where quasi-TE and quasi-TM modes of x-cut lithium niobate can hybridize in waveguide bends[6,38]. Lithium tantalate's low birefringence relaxes this constraint, allowing waveguide dimensions to be optimized for dispersion and confinement without compromising mode purity. Combined with the coupling control demonstrated here, this points toward resonant second-order devices that are simultaneously broadband and optimally coupled. Together with high damage threshold, low photorefraction, and broad transparency window, these results establish the LTOI platform for efficient, low-power frequency conversion and open a route to squeezed-light generation, photon-pair sources, and telecom-to-visible quantum interfaces on a single chip.

## Methods

### Numerical simulations

The bus-ring coupler was simulated with the finite-difference time-domain (FDTD) method using Tidy3D. The mode profiles at the fundamental and second-harmonic wavelengths (Fig. 1b) and the per-pass coupling were obtained from a single simulation of the coupling region, in which the tapered bus and the racetrack straight run parallel over the coupler length. A guided mode was launched in the bus ($TE_{00}$) using the built-in mode solver, and mode monitors on the bus (through) and ring (drop) ports were projected onto the $TE_{00}$ modes of each waveguide to give the per-pass power coupling coefficient $K_{\mathrm{pass}}$ as a function of wavelength over the 0.6 to 1.7 µm range.

The per-pass coupling was converted to a coupling rate and extrinsic quality factor using $Q_e = \frac{\omega}{\kappa_e} = \frac{2\pi\nu}{K_{\mathrm{pass}}\,\mathrm{FSR}}$, with FSR the measured free spectral range and $\nu$ the frequency [Hz]. The extraction efficiency $\eta = \frac{Q_i}{Q_i+Q_e}$ was then evaluated assuming an intrinsic quality factor $Q_i$= 2×10$^6$ in both bands, consistent with the measured values. The coupler length L and gap were swept to map $\eta$ at the fundamental and second-harmonic bands (Fig. 1c).

**Device fabrication**

Devices were fabricated on an x-cut thin-film lithium tantalate on insulator wafer with a 500 nm LT device layer on $SiO_2$. The optical layer, consisting of poling calibration waveguides and racetracks, was patterned by electron-beam lithography using ma-N 2405 negative resist (e-beam current: 1nA, dose 700 uC/cm2) , and the pattern was transferred into the lithium tantalate by argon-based reactive-ion etching to an etch depth of ~300 nm, leaving a partially etched slab. After etching, the redeposition (etch byproduct along the waveguide sidewalls) was removed with a wet etch and verified by SEM. The chip is annealed in oxygen at 520 °C.

The chip was diced, separating the calibration waveguides and the racetrack devices into two chiplets. The calibration chiplet was processed first. The poling-finger electrodes were written in ZEP520A by electron-beam lithography aligned to the optical layer using local markers, with the electrodes placed on the sidewalls of the waveguide (sidewall poling). Chromium/gold electrodes (Cr/Au, 15/125 nm) were deposited by electron-beam evaporation and patterned by lift-off in PG remover. Periodic domain inversion was achieved by applying high-voltage pulses to the electrodes (peak 300 V, 2 ms duration), with the poling monitored by second-harmonic microscopy. Wavelength-calibrated adaptive poling was used: the local poling period on the racetracks chiplet was set from an interpolated map built from ellipsometry (film thickness) and mode-solver data sampled along the resonator top straight, correcting for thickness variation across the device and adjusting for a global offset extracted from SHG measurement of the calibration chiplet. The same poling procedure is then finally performed on the racetracks. The waveguide facets were prepared by dicing and polishing for fiber coupling.

**Experimental setup**

We characterize the fabricated devices in a fiber-to-fiber transmission setup using two different lasers and paths to couple light into the chip and one common output path.

In the SH input path, we use HI 780 fibers. The laser (New Focus Velocity TLB-6700) output is split (99/1%) after an optical isolator. The 1% tap goes to an imbalanced MZI (1 m imbalance, FSR=192 MHz) for relative frequency calibration, while the 99% goes through an electronic variable optical attenuator (Thorlabs V800A) and through a manual fiber polarization controller (Thorlabs FPC030). Next, another 1% is tapped out to a Si amplified detector (Thorlabs PDA36A2) for input power monitoring while the rest of the light is coupled into the chips with a visible single-mode lensed fiber (OZ Optics).

The telecom (FH) input path uses SMF-28 fibers. After polarization control (Thorlabs FPC560), 1% of the laser output (Santec TSL-570) is sent to an InGaAs amplified detector (Thorlabs PDA20CS2) for input power monitoring, and the rest is coupled to the chip with a telecom single-mode lensed fiber (OZ Optics). The chip is mounted on a custom-made stage whose temperature is regulated with a TEC controller (SRS LDC502).

The output path is the same for both input paths except the output lensed fiber, which is swapped depending on the measurement performed. The output collected from the lensed fiber is spectrally split by a WDM (1560nm/780nm PM Optics) into a visible and a telecom output path. The telecom path goes through an isolator that is opaque to visible light and is then split 99/1% between an InGaAs APD (Thorlabs APD431C) and a power meter (Thorlabs PM100D, S155C). The visible path is split 90/10% between an Si amplified detector (Thorlabs PDA100A2) and a power meter (Thorlabs PM100D, S150C). For the spectrally resolved measurements of the OPO we add an extra tap to the telecom output path and send it to an optical spectrum analyzer (Yokogawa AQ6370D). All the splitters have been independently measured to extract the experimental splitting ratio as a function of the wavelength. The experiment is controlled with a data acquisition system (DAQ, National Instruments USB-6346). All the detectors were connected to analog input ports, while the visible laser piezo and VOA were controlled via the analog outputs.

**Modeling**

We analyze both devices within the standard input-output framework for doubly resonant $\chi^{(2)}$ microresonators, in which the interacting modes obey coupled-mode equations of motion and the nonlinearity enters through a single-photon coupling rate *g*. The derivation and notation adopted here follow the treatment of ref.[22]

OPO threshold and nonlinear coupling

The OPO output power was fit to the above-threshold law of Eq. 2, with all intrinsic and extrinsic rates fixed by fitting independent low-power linear transmission scans using the standard coupled-mode theory lineshape. The fitted parameters are $P_{th}$ and the output facet loss; input facet loss at SH is fixed by independent measurement to 6.7 dB. The fit was restricted to a single OPO pair (signal: $\lambda_s = 1545.96$, $\kappa^{(e)}_{A,+m}/2\pi = 30.42$ MHz, $\kappa^{(i)}_{A,+m}/2\pi = 85.69$ MHz, $Q_e$= 6.37×10$^6$, $Q_i$= 2.26×10$^6$, $Q_{load}$=1.67×10$^6$, $\eta_{(A,m)}$= 0.26; idler: $\lambda_i = 1570.71$, $\kappa^{(e)}_{A,-m}/2\pi = 19.18$ MHz; $\kappa^{(i)}_{A,-m}/2\pi = 90.63$ MHz, $Q_e$ = 9.95×10$^6$, $Q_i$ = 2.11×10$^6$, $Q_{load}$ = 1.74×10$^6$, $\eta_{(A,-m)}$= 0.17; SH pump $\lambda_p$ = 779.14 nm; $\kappa^{(e)}_B/2\pi$=160.06 MHz, $\kappa^{(i)}_B/2\pi$=199.11 MHz, $Q_e$ = 2.40×10$^6$, $Q_i$= 1.93×10$^6$; $Q_{load}$ = 1.07×10$^6$, $\eta_B$= 0.45). This gives $P_{th} =$ 180.98 μW and 4.1 dB facet loss, which we confirmed with an independent fiber-to-fiber loss measurement. The corresponding effective nonlinear coupling is $g_{eff}/2\pi$ = 39.6 kHz, under the assumption that $(\Delta + \mu - D_2 m^2) \approx 0$. The theoretical value calculated via mode solver for the same configuration is $g_{0,-mm}/2\pi = 63$ kHz.

SHG efficiency model

For the SHG device, the fundamental (FH) is the pump and the second harmonic (SH) is generated, so the roles of the two modes are reversed relative to the OPO. From the steady-state solution of the SH and FH equations of motion (including FH drive), it is possible to obtain the implicit (cubic) relation that the intracavity FH photon number $n_A$ has to satisfy

$$n_A \left| i\delta - \frac{\kappa_A}{2} + \frac{2|g|^2 n_A}{i(2\delta - \mu) - \frac{\kappa_B}{2}} \right|^2 = \frac{\kappa_A^{(e)} P_{in}}{\hbar\omega_0}$$

Where $\delta = \omega_L - \omega_0$ is the FH pump detuning. The corresponding fundamental transmission lineshape, including these nonlinear terms, is

$$T(\delta) = \left| 1 + \frac{\kappa_A^{(e)}}{i\delta - \frac{\kappa_A}{2} + \frac{2|g|^2 n_A}{i(2\delta - \mu) - \frac{\kappa_B}{2}}} \right|^2$$

and the generated second-harmonic power at the output is $P_{SH} = \frac{\hbar\omega_B \kappa_B^{(e)} |g|^2 n_A^2}{\left|i(2\delta-\mu)-\frac{\kappa_B}{2}\right|^2}$, giving an on-chip conversion efficiency $\eta_{SH} = \frac{P_{SH}}{P_{in}}$.

Cavity rates were obtained from linear transmission fits as described above. At the resonance used here ($\lambda_{\mathrm{FH}}$ =1559.89 nm), the fundamental has $\mathrm{Q_i}$ = 2.0×10$^6$ and $\mathrm{Q_e}$= 3.9×10$^5$ ($\kappa_A^{(e)}/2\pi = 490.35$ MHz, $\kappa_A^{(i)}/2\pi = 96.32$ MHz), hence a loaded $\mathrm{Q_{load}}$ = 3.3×10$^5$ and an extraction efficiency $\eta_A$ = 0.84, i.e., strongly overcoupled by design. The second-harmonic mode at 779.94 nm has $\mathrm{Q_i}$=2.1×10$^6$ and $\mathrm{Q_e}$ = 3.2×10$^6$ ($\kappa_B^{(e)}/2\pi = 119.31$MHz, $\kappa_B^{(i)}/2\pi = 186.98$ MHz), giving $\mathrm{Q_{load}}$ = 1.25×10$^6$ and $\eta_B$ = 0.39.

The nonlinear coupling rate was determined by fitting the power-dependent fundamental transmission lineshapes to the model above, solving the cubic relation for $n_A$ at each detuning and power. A single global $\mathrm{g_{eff}}$ was used, together with a per-scan fundamental resonance offset $\delta_0$ (accounting for thermal drift) and SH detuning μ. Only the line-scans that do not show cascaded OPO were used for the fit. This yields $\mathrm{g_{eff}}/2\pi$ = 60.5 kHz, close to the numerically calculated value $\mathrm{g_{0,00}}/2\pi = 63$ kHz. The efficiency curve of Fig. 5e was then computed on resonance from these independently determined parameters, with the input facet loss fixed at 4.19 dB and the output-coupling loss as the only adjusted parameter, giving 5.45 dB, in agreement with an independent measurement. It is worth pointing out that the conversion efficiency can be pushed even higher by intentionally overcoupling the visible: the fundamental should reach critical coupling once the nonlinear loss is included, while the second harmonic is correspondingly overcoupled so that the generated light is extracted as efficiently as it is produced[23]. Because the coupler sets the coupling in each band independently, this joint condition is within reach and points toward substantially higher on-chip conversion.

**Acknowledgments**

The authors thank T. McKenna for insightful discussions and A. McClelland for technical support on SH microscope imaging. The device fabrication in this work was performed at the Harvard University Center for Nanoscale Systems, a member of the National Nanotechnology Coordinated Infrastructure Network, which is supported by the US National Science Foundation under NSF award no. ECCS-2025158.

Funding: Rubicon postdoctoral fellowship from the Dutch Research Council (019.231EN.011), Gordon and Betty Moore Foundation (A.C.), NSF Center for Quantum Networks (ERC EEC-1941583) (J.Y., S. L.), Air Force Office of Scientific Research FA955025CB011, Naval Sea Systems Command N0002425CT146 (Y.S.), Harvard Grid Fellowship (K.P.), National Research Foundation of Korea (X.L. T.G.), Amazon Web Services - A60290 (D.W.), Monarch Quantum, Inc. - A61787 (I. O., N. A., A. S.), Defense Advanced Research Projects Agency HR0011-24-2-0360 (X.Z.)

**Author contributions**

AC conceived the device design through numerical simulations and theoretical analyses, supported by YS, TG, and NA. AC fabricated and characterized the devices with the support of all co-authors. ML supervised the project. All authors contributed to the analysis and writing of the paper.

**Competing interests**

N.S., K.P. and M.L. are involved in developing lithium tantalate technologies at Lumina Corporation. The remaining authors declare no competing interests.

**Data availability**

The data that support the findings of this study are available from the corresponding author upon reasonable request.

**Code availability**

All codes produced during this research are available from the corresponding author upon reasonable request.

**ORCID**

A. Cordaro: 0000-0003-3000-7943

Y. Song: 0000-0003-3879-4777

K. Powell: 0000-0001-6327-1893

X. Li: 0009-0006-1186-050X

T. Ge: 0009-0008-7560-0250

X. Zhu: 0009-0000-3194-0693

I. Occhiodori: 0009-0007-2249-8355

S. Lu: 0009-0000-9318-9815

J. Yang: 0000-0001-5102-1002

M. Lončar: 0000-0002-5029-5017